%% file: main.tex
\documentclass{MITcsail}
\usepackage{graphicx}
\usepackage{comment}
\input{includes/definitions}

\title{
Robust resonant anomaly detection with NPLM
}
\author
{
Gaia Grosso~$^{1, 2, 3,}$\footnote{\href{mailto:gaiag795@mit.edu}{gaiag795@mit.edu}}, 
Debajyoti Sengupta~$^{4,}$\footnote{ \href{mailto:mdebajyoti.sengupta@unige.ch}{debajyoti.sengupta@unige.ch}}\footnote{Work done as a Ph.D. student at the University of Geneva}, 
Tobias Golling~$^{4}$, and 
Philip Harris~$^{1, 2}$\\
\vspace{1em} 
\normalfont{\small $^{1}$NSF AI Institute for Artificial Intelligence and Fundamental Interactions, Cambridge, MA}\\
\normalfont{\small $^{2}$MIT Laboratory for Nuclear Science, Cambridge, MA}\\
\normalfont{\small $^{3}$School of Engineering and Applied Sciences, Harvard University, Cambridge, MA}\\
\normalfont{\small $^{4}$Département de Physique Nucleaire et Corpuscolaire, University of Geneva, Geneva, Switzerland}\\
}

\begin{document}
\maketitle
\thispagestyle{firstpagestyle}
    \begin{abstract}
       In this study, we investigate the application of the New Physics Learning Machine (NPLM) algorithm as an alternative to the standard \cwola method with Boosted Decision Trees (BDTs), particularly for scenarios with rare signal events. NPLM offers an end-to-end approach to anomaly detection and hypothesis testing by utilizing an in-sample evaluation of a binary classifier to estimate a log-density ratio, which can improve detection performance without prior assumptions on the signal model. We examine two approaches: (1) a end-to-end NPLM application in cases with reliable background modelling and (2) an NPLM-based classifier used for signal selection when accurate background modelling is unavailable, with subsequent performance enhancement through a hyper-test on multiple values of the selection threshold. Our findings show that NPLM-based methods outperform BDT-based approaches in detection performance, particularly in low signal injection scenarios, while significantly reducing epistemic variance due to hyperparameter choices. This work highlights the potential of NPLM for robust resonant anomaly detection in particle physics, setting a foundation for future methods that enhance sensitivity and consistency under signal variability.
    \end{abstract}

    \section{Introduction}

\input{tex/intro}
    
    \section{Classifier based resonant anomaly detection}
    \input{tex/cwola}
    \section{The New Physics Learning Machine (NPLM)}
    \input{tex/nplm}

    \section{Numerical experiments}\label{sec:experiments}
    \input{tex/data}
    \input{tex/results}

    \section{Summary and outlook}
    \input{tex/outlook}

    \section*{Acknowledgements}
    \input{includes/acknowledgement}

    \bibliographystyle{hunsrt}

\input{main.bbl}
\end{document}

%% file: includes/definitions.tex
\usepackage{xspace}

\newcommand\cwola{CWoLa\xspace}

\newcommand{\mjj}{\ensuremath{m_{JJ}}\xspace}

\def\data{{\cal{D}}}

\def\Ndata{{{\rm{N}}_{\cal{D}}}}
\def\reference{{\cal{R}}}

\def\Ndata{{\rm{N}}_{\cal{D}}}

\newcommand\RH{{\rm{R}}}

\newcommand\w{{\rm{\bf{w}}}}

\newcommand\HH{{\rm{H}}_{\rm{\bf{w}}}}

\newcommand\Hnull{{\rm{H_{\boldsymbol0}}}}

%% file: tex/intro.tex

Detecting narrow peaks in smoothly declining background distributions is essential in collider data analysis.
This is because new particles or massive mediators appear as resonant peaks in the invariant mass spectrum, with widths correlating to their decay times. Such ``bumps" are hints of a potential new particle production. Resonant searches facilitate data-driven, signal-agnostic exploration, where background estimates are generated by smoothly fitting sideband data, allowing for flexible, even simple, signal hypotheses in the region of interest.

Over time, statistical methods for resonance searches have evolved, improving background estimation and signal modelling. Recently, machine learning has been integrated into resonance detection, enhancing data utilization while preserving signal independence~\cite{anode,cathode,LaCathode,curtains,curtainsf4f,drapes,ranode,conrad}. These methods, named resonant anomaly detection, have significantly increased sensitivity to new physics signals in ATLAS and CMS analyses~\cite{CMS:2024lwn,ATLAS:2020iwa}.

ML enhances resonance searches across three stages: (1) background template generation in signal regions via generative modelling of sidebands, (2) anomaly detection for signal enhancement, and (3) hypothesis testing through signal-agnostic goodness-of-fit approaches.
In this work, we focus on the second and third steps, namely the definition of a signal-enriched selection and the hypothesis test. 
We review the state-of-the-art machine learning methods used to address the problem of anomaly detection based on Boosted Decision Trees (BDTs) as binary classifiers and compare it with the New Physics Learning Machine (NPLM) algorithm~\cite{nplm}. 

The NPLM algorithm is an end-to-end approach 
to simultaneously detect and test for the presence of anomalies, without prior assumptions on the signal model. NPLM relies on the \textit{in-sample} evaluation of a binary classifier to estimate a log-density-ratio between experimental data and a reference sample, and computes the Neyman-Pearson likelihood-ratio test according to it.
The in-sample nature of the procedure has the potential to improve on current methods based on standard out-of-sample binary classification in scenarios of extremely rare signals.
At the same time, the NPLM algorithm requires a highly accurate background template, which is often not available.

Hence, we study two different solutions to resonant searches based on NPLM: (1) in presence of a good background modelling we skip the selection stage and run the NPLM algorithm as an end-to-end approach; (2) instead, 
in absence of a good background modelling, we replace 
the standard BDT-based classifier 
with an NPLM-classifier in the anomaly selection stage. 

Our study represents a first example of how the NPLM strategy can be integrated in the classic bump-hunt approach based on a sliding window.

Compared to current BDT-based approaches, we find that the NPLM-based approaches reach better detection performances with significantly reduced epistemic variance in low signal injection scenarios, that are particularly relevant for realistic applications. 

%% file: tex/cwola.tex
\label{sec:classifierad}
Machine learning (ML) is widely used for resonant anomaly detection in high-energy physics (HEP), enhancing the identification of rare events and potential new physics within massive datasets from particle accelerators like the LHC. Specifically, in resonance detection, ML helps isolate anomaly-enriched data subsets for further hypothesis testing with classic bump-hunt methods. ML-based anomaly detection typically falls into three categories: (a) weakly-supervised methods, which use noisy labels to separate anomalous data from expected background, (b) semi-supervised methods, which aim to use partially labeled datasets, and (c) unsupervised methods, such as autoencoders and clustering, which detect rare or out-of-distribution data without labelled input.


\cwola~\cite{cwola,cwolabump} underpins various weakly supervised anomaly detection algorithms. The Neyman-Pearson lemma states that the most powerful test statistic for testing a data sample $\data$ originated from a signal hypothesis $\rm S$ against a background hypothesis $\rm B$ is the log-likelihood ratio, $t(\data) = 2\log\frac{{\cal L}(\data|{\rm S})}{{\cal L}(\data|{\rm B})}$. For selecting anomaly-enriched regions, this criterion is used to score data points based on the density ratio or its logarithmic form.
\begin{equation}\label{eq:class-score}
c(x)=\log\frac{n(x|{\rm S})}{n(x|{\rm B})}
\end{equation}
and retain only those exceeding a predefined threshold ($\rm thr$).

With a specified signal hypothesis, a fully supervised binary classification task can distinguish between signal and background events. However, without a signal hypothesis, this approach is unfeasible due to the absence of labels in real data. Assuming a localized signal, a \emph{signal region} (SR) can be defined in the data space where the signal is relatively enriched compared to surrounding \emph{sidebands} (SB). \cwola demonstrates that the density ratio from classifying SB versus SR can approximate Eq.~\ref{eq:class-score}.

However, if the features used to train the classifier are correlated with the feature used to define the SR, the classifier will learn the differences between SR and SB due to the ``sculpting" introduced by the region condition, even if a signal is not present.
To circumvent this, several methods have been proposed to construct a background \emph{template} (e.g. a reference model) in the SR from the data.
The classifier is then trained to distinguish between the template and the SR. 
To single out the impact of the anomaly selection step in the resonance search, in this work we will assume the existence of a perfect template in the SR, or the so called idealised setting as shown in~\cite{anode,curtains,curtainsf4f}.

\subsection{Boosted Decision Trees}
Typically, deep neural networks are employed as classifiers in the template based weakly supervised methods.
More recently, Boosted Decision Trees (BDTs) have surfaced as a more effective means of anomaly detection~\cite{backtoroots}, and have been shown to be more sensitive at lower signal strengths compared to MLP based classifiers.
Furthermore, BDTs are relatively unaffected by uninformative features, which is valuable in high-dimensional spaces which may contain irrelevant data for a given signal model.
This robustness, combined with efficient training and evaluation, make BDTs a compelling alternative to deep neural networks for model-agnostic anomaly detection.
The key hyperparameters characterizing the BDTs are:
\begin{enumerate}
    \item $i_{\rm max}$: the maximum number of iterations of the boosting process, or the maximum number of trees for binary classification;
    \item $n_{\textrm{leaf}}$: the maximum number of leaf nodes for each tree;
    \item $\eta$: the learning rate for the trees;
    \item $\lambda$: the L2 regularisation parameter for leaves with small hessian values.
\end{enumerate}

The maximum number of leaf nodes ($n_{\textrm{leaf}}$) controls the model's complexity, with higher values allowing each tree to capture more detailed patterns, potentially increasing accuracy but also raising the risk of overfitting.
The maximum number of iterations ($i_{\rm max}$) dictates how many trees are sequentially added to correct previous errors, with more iterations generally improving performance up to a point where additional iterations may lead to overfitting and diminishing returns.
The L2 regularization term, whose impact is determined by $\lambda)$, helps to prevent overfitting by penalizing large weights and thus encouraging the simpler solutions that generalize better to new data.
The learning rate $\eta$ controls how much each new tree contributes to the ensemble, where smaller rates slow down the learning process but often improve stability and accuracy, as long as a sufficient number of iterations is used to capture complex patterns.
Finally, a crucial hyperparameter is the choice of the background rejection threshold $\rm thr$, which is applied to the data to select the `most anomalous samples'.

\paragraph{Implementation of BDT-based classifier for anomaly detection}
The BDTs in this study are implemented using \texttt{scikit-learn} package.
To ensure robustness against overfitting, the BDTs are trained in a kFold cross validation.
KFold is a technique used to evaluate the performance of a machine learning model by dividing the data into k equal-sized folds. Each fold is used as a validation set while the remaining k-1 folds are used for training, and this process is repeated k times to ensure that every data point is used for both training and validation. This method helps to reduce overfitting and provides a more accurate estimate of the model's performance on unseen data.
An ensemble of 50 Histogram-Gradient BDTs are trained with kFold cross validation with k = 5.

%% file: tex/nplm.tex
New Physics Learning Machine (NPLM) offers an alternative to standard classifiers for anomaly detection that returns both the equivalent to classifier scores and a final test statistic, readily available to compute a discovery $p$-value.
The NPLM algorithm~\cite{DAgnolo:2018cun,DAgnolo:2019vbw,Letizia:2022xbe} is based on the classical likelihood ratio hypothesis testing method introduced by Neyman and Pearson~\cite{Neyman:1933wgr}. Given a null hypothesis, $\Hnull$, and a family of alternatives, $\HH$,  the NPLM method aims at computing the test statistic defined as twice the maximum log-likelihood-ratio over the space of the alternatives: 

\begin{equation}\label{eq
}
t(\data) = 2 \max\limits_\w \sum\limits_{x\in \data} \log\frac{{\cal L}(\data|{\HH})}{{\cal L}(\data|\Hnull)}.
\end{equation}

A model $f_\w(x)$ with trainable parameters $\w$ is used to parameterize a family 
$\HH$ of alternative hypotheses for the data density:
\begin{equation}\label{eq:n}
n(x|\HH) = n(x|\RH) \exp[f_\w(x)]
\end{equation}

This approach allows the maximum-likelihood problem to be addressed in a signal-independent manner by leveraging data evidence. The problem is reframed as a machine learning task with the following loss function:

\begin{equation}\label{eq:l}
L_{\rm NPLM}[f_{\w}] = \sum\limits_{x \in \reference} w_x (e^{f_\w} - 1) - \sum\limits_{x \in \data} f_\w .
\end{equation}

Here, $\data$ is the sample of interest, and $\reference$ is a reference dataset representing the nominal condition under the 
$\RH$ hypothesis. It should be noted that the reference sample size has to exceed $\Ndata$ to ensure robustness against statistical fluctuations, and to overcome the sample imbalance, the weights $w_x$ have to be adjusted to match the expected number of data points under nominal conditions N(R).

The test statistic, indicating the degree of abnormality in $\data$, is computed as minus twice the value of the loss function at the end of the minimization routine:

\begin{equation}\label{eq:t}
t_{\rm NPLM}(\data) = -2 \min\limits_{\w} L_{\rm NPLM}[f_{\w}]
\end{equation}

Calibration of the NPLM test involves empirically estimating the distribution 
$p(t_{\rm NPLM}|\RH)$ by conducting pseudo-experiments with artificial datasets generated under the $\RH$ hypothesis. The test value for 
$\data$ is compared to this distribution to compute the 
$p$-value:
\begin{equation}\label{eq:p}
p\text{-value} = \frac{1}{|{\rm T_R}|} \sum_{t \in {\rm T_R}} {\rm I}[t > t(\data)]
\end{equation}

\subsection{NPLM for local anomaly detection}
NPLM offers a general framework to test the compatibility of a dataset $\data$ with a distribution of reference. The lack of assumptions on the nature of the anomalous signal make it adaptable to various searching scenarios, resonant signals included. The only requirement to run NPLM besides the experimental data is the existence of a highly accurate reference model to represent the data behaviour in absence of signal. If such a model existed for the full phase space of the analysis and was not affected by systematic uncertainties, then NPLM could be run over the inclusive sample at once, and a global $p$-value could be straightforward computed as in Eq.~\ref{eq:p}. However, in real applications at collider experiments, such reference model is often not well known. There are situations in which the lack of information about the reference model can be embodied in a set of nuisance parameters. In that case, an extended version of NPLM that includes the treatment of systematic uncertainties can be performed (see ~\cite{dAgnolo:2021aun} for details). On the other hand, in many other cases the only way to properly build a reference model is data-driven: some control regions assumed signal-free are fitted with a smooth function, and the result is then transferred to the signal region to constitute the background template.
Data-driven ways to construct the background model are often preferred over simulation based ones because they can be more reliable and almost free of systematic uncertainties. However, they require a strong assumption on the signal location.
In the case of resonant searches, the assumption of a narrow over-density as a signal allows for the use of sidebands around the resonance as control regions from which to extract a background modelling. In this case, the resonant variable is scanned with sliding signal regions, and for each step in the scan a local test over the signal region is performed. In this scenario, NPLM can be used to perform the local test, and the global $p$-value can then be obtained by accounting for the look elsewhere effect associated to the multiple testing.

Two distinct approaches could be taken in resonant searches:
\begin{itemize}
    \item \textbf{NPLM-classifier}: the NPLM model is used to replace the standard classifier and perform anomaly selection based on the score. A cut-and-count test is performed downstream. 
    \item \textbf{NPLM-end-to-end}: the NPLM model is run over the full set of variables, the auxiliary ones as well as the resonant one, without imposing any selection.
\end{itemize}


In Section~\ref{sec:experiments} we compare the two approaches numerically using the LHCO dataset as a benchmark.

\subsection{NPLM implementation}
The NPLM algorithm has been originally implemented using dense neural networks~\cite{DAgnolo:2019vbw}, and kernel methods later~\cite{Letizia:2022xbe}. The latter implementation relies on the \textsc{Falkon} library~\cite{rudi2017falkon}, a framework to solve kernel ridge regression problems in a time efficient way, optimizing the computation on GPUs~\cite{meanti2020kernel}.
Both the neural network and the kernel methods based models are characterized by a set of hyperparameters.
For the neural networks, those are mainly the ones defining the architecture of the model (e.g. number of layers and number of nodes per layer), and the weight clipping parameter, defining the regularization scheme.
Ref.~\cite{DAgnolo:2019vbw} introduced a heuristic approach solely based on background-like data to select suitable models, recovering good statistical properties of the NPLM test statistic distribution in the null hypothesis (e.g. asymptotic $\chi^2$ behaviour).
Briefly, Ref.~\cite{DAgnolo:2019vbw} shows that the weight clipping parameter can be tuned studying the test statistic distribution under the null, and an optimal value can be selected that brings the distribution close to a $\chi^2$ with number of degrees of freedom equal to the number of trainable parameter of the model.
A family of well behaving models can be selected in this way, and additional studies presented in Ref.~\cite{Grosso:2023hew} show no major difference in performances within the selected models.

In this work, we will focus on the kernel methods implementation because it is significantly faster than the one based on neural networks.
The kernel methods implementation based on \textsc{Falkon} is characterized by three main hyperparameters: 
\begin{itemize}
    \item $M$: the number of kernels;
    \item $\sigma$: the gaussian kernels' width;
    \item $\lambda$: the L2 regularization coefficient.
\end{itemize}

A heuristic procedure to select reasonable candidates for $M$, $\sigma$ and $\lambda$ was proposed in Ref.~\cite{Letizia:2022xbe} and, recently, new studies have showed a way to use multiple testing to avoid the selection of $\sigma$~\cite{Grosso:2024wjt}.
%

%% file: tex/data.tex
To easily compare existing approaches to resonant anomaly detection with NPLM we consider as physics case the search for anomalies in a dijet final state, a widely used benchmark in the community.
\subsection{Datasets}
We consider the following two datasets.
\begin{itemize}
\item\textbf{LHCO dataset.}
The LHCO R\&D dataset~\cite{LHCOlympics} comprises background data produced through QCD dijet production,
with signal events arising from the all-hadronic decay of a massive particle to two other massive particles
$W^\prime\rightarrow X(\rightarrow q\bar{q}) Y(\rightarrow q\bar{q})$, each
with masses $m_{W^\prime} = 3.5$~TeV, $m_{X} = 500$~GeV, and $m_{Y} = 100$~GeV.
Both processes are simulated with \texttt{Pythia}~8.219~\cite{pythia} and interfaced to \texttt{Delphes}~3.4.1~\cite{deFavereau:2013fsa} for detector simulation.
Jets are reconstructed using the anti-$k_\mathrm{T}$ clustering algorithm~\cite{Cacciari:2008gp} with a radius parameter $R=1.0$, using the \texttt{FastJet}~\cite{Cacciari:2011ma} package. 
Events are required to have at least one jet with pseudorapidity $\left| \eta \right| < 2.5 $, and transverse momentum $p_\mathrm{T}^{J} > 1.2$~TeV.
The top two leading $p_\mathrm{T}$ jets are selected and ordered by decreasing mass.
In order to remove the turn on in the \mjj distribution arising from the jet selections, we only consider events with $\mjj > 2.8$~TeV.
To construct the training datasets, we use varying amounts of signal events mixed in with the QCD dijet data.
In total there are 1~million QCD dijet events and, 100\, 000 signal events.

\item\textbf{RODEM dataset.}
The RODEM dataset~\cite{rodemjets} consists of jet samples originating from multiple sources.
We use a type-II two-Higgs-doublet model (2HDM) generated with the FeynRules package~\cite{feynrules} (v2.3.24).
Specifically, the process $pp \rightarrow H^0 \rightarrow h^+ h^-$, where H denotes a heavy neutral Higgs scalar and $h^{\pm}$ a lighter charged Higgs scalar, is considered, where $h$ decays as $h \rightarrow t b$ leading to a 4-pronged jet substructures.
For this signal, a different event selection is applied.
Reconstructed events are required to have a jet with transverse momentum $p_T > 450$ GeV.
No additional event selection criteria are imposed, and additional jets are not vetoed.
A total of 100\,000 events are simulated using $m_H = 2250$ GeV and $m_h = 500$ GeV.\\
\end{itemize}

To study the performance of our method in enhancing the sensitivity in a bump hunt, we use the six input features proposed in Refs.~\cite{cwolabump,anode,cathode,curtains,curtainsf4f}:
the dijet invariant mass ($m_{JJ}$) as the resonant variable, and five additional auxiliary variables listed below
\begin{itemize}
    \item $m_{J_1}$: the invariant mass of the most energetic jet;
    \item $\Delta m_{J}$: the relative distance between the two jets' masses;
    \item $\tau_{21}^{J_1}$: the ratio between the 1 and 2-subjettinness~\cite{nsubjettiness} of the first jet;
    \item $\tau_{21}^{J_2}$: the ratio between the 1 and 2-subjettinness of the second jet;
    \item $\Delta R_{JJ}$: the angular separation between the two jets, computed as the Euclidean distance in the $\eta-\phi$ plane: $\Delta R_{JJ}= \sqrt{\Delta \eta_{J_1,J_2}^2+\Delta\phi_{J_1,J_2}^2}$
\end{itemize}

For both datasets, a fixed signal region of 3300 - 3700 GeV is considered, with expected background yield of $121339$.
In this SR, the LHCO signal forms a localised resonance, whereas the 2HDM signal is a flat excess in the invariant mass.
This effectively mimics non-resonant signals resulting from off-shell effects or final states with large missing energies.

%% file: tex/results.tex
\begin{figure}[t!]
    \centering    \includegraphics[width=\linewidth]{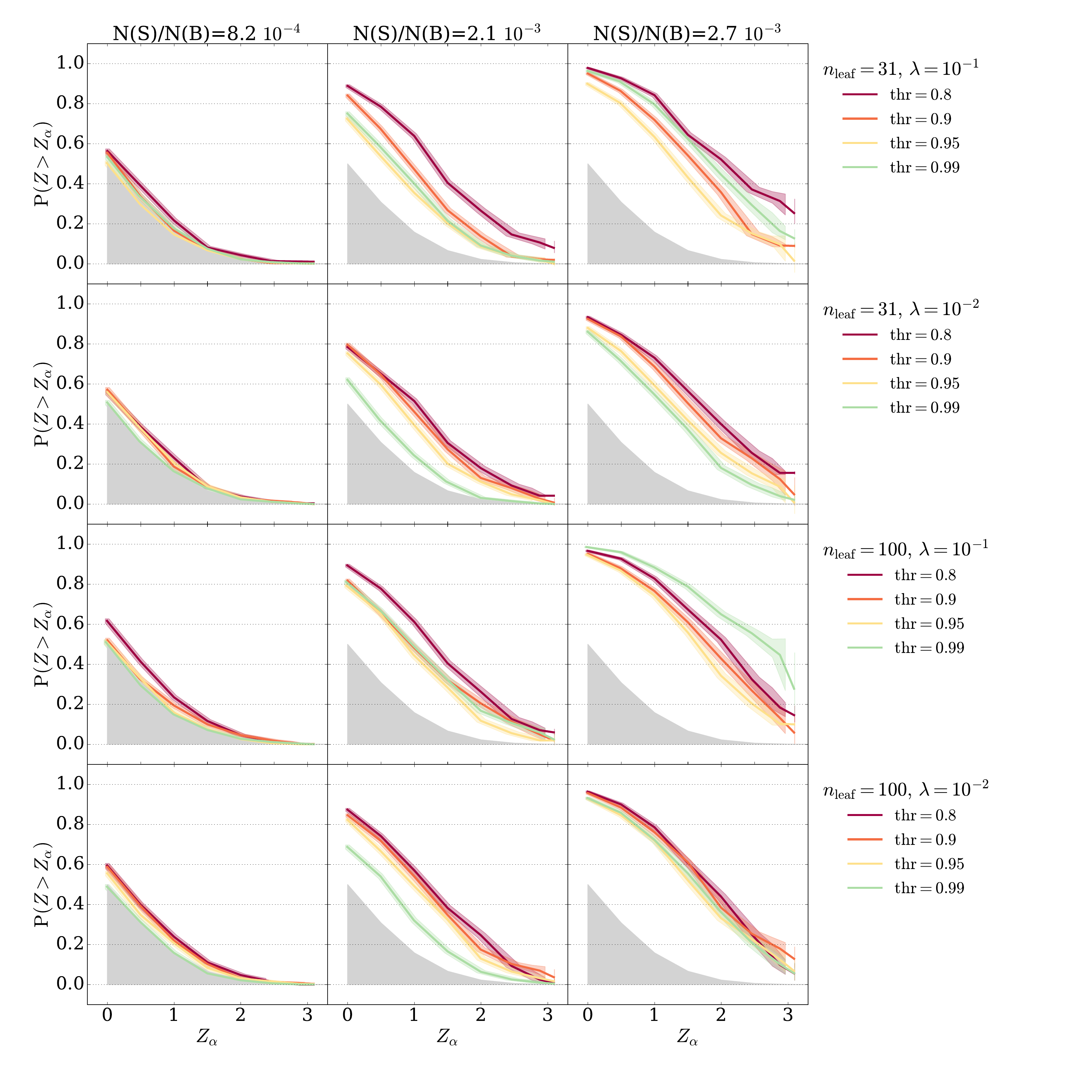}
    \caption{\textbf{BDT-classifiers: response across different hyperparameters choices.} Discovery power of the cut-and-count signal-agnostic test on the score of a BDT classifier. Different colours represent different BDT hyperparameters choices. The observed variance within each row is given by the choice of the anomaly selection threshold ($\rm thr$) applied on the score of the classifier before performing the test. We consider four values for $\rm thr$: 0.8, 0.9, 0.95, 0.99.}
    \label{fig:bdt-all}
\end{figure}
\subsection{Sensitivity of BDT classifiers to hyperparameters selection}
The first study we perform aims at highlighting the limitations of implementations based on BDT classifiers for anomaly detection. The limitations we observed are mainly related to the stability of the algorithm response across hyperparameters choices. 

We run various BDT models with different hyperparameter choices, and we observe the detection ability after an anomaly selection threshold ($\rm thr$) is applied. This is essentially a background rejection factor. For simplicity, we run the comparison on a simple cut-and-count ($\rm cc$) statistic computed over the events surviving the selection:
\begin{equation}
    t_{\rm cc}(\data) = 
    \frac{N(x\in \data |c(x)>{\rm thr} ) - N(x\in \reference|c(x)>{\rm thr})}{\sqrt{N(x\in \reference|c(x)>{\rm thr})}}
\end{equation}
where $\data$ and $\reference$ refer to data and background template in the SR.
As mentioned before, $\reference$ in this work pertains to the idealised setting.

To calibrate the test statistic, we repeat the analysis with signal-free toys and recompute $t_{\rm cc}$. 
Our results are presented in Figure~\ref{fig:bdt-all} in terms of power curves, showing the probability of observing a $p$-value smaller or equal to a false positive threshold $\alpha$ as a function of the associated $Z$-score, $Z_{\alpha}$. 
The latter is defined as the quantile $\Phi^{-1}$ of the normal distribution at the $\alpha$ complement to 1
\begin{equation}
    Z_{\alpha} = \Phi^{-1}(1-\alpha).
\end{equation}
The gray filled area in the plots represents the power in absence of signal and serves as a baseline.

\begin{figure}[t!]
    \centering
    \includegraphics[width=\linewidth]{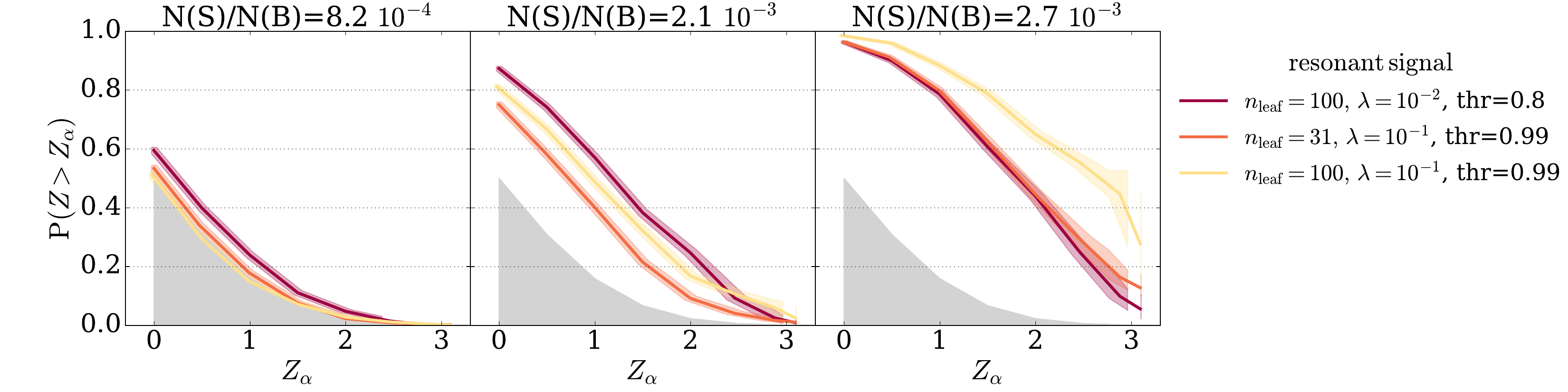}
    \includegraphics[width=\linewidth]{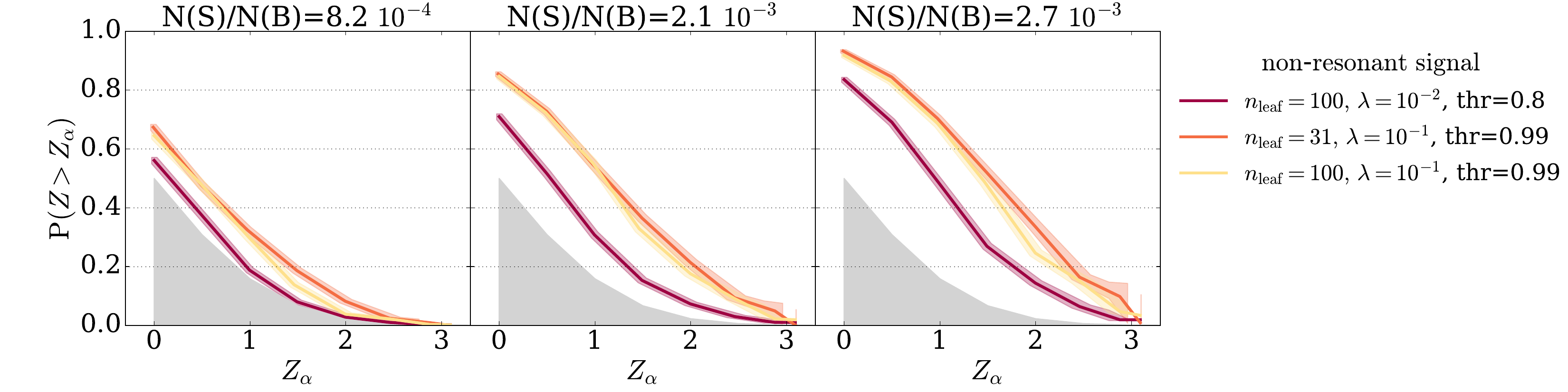}
    \caption{\textbf{BDT-classifiers: response to different signal models.} Discovery power of three BDT-based classifiers for the resonant signal in the LHCO dataset (top row) and the non-resonant signal in the RODEM dataset (bottom row). Detection performances highly depend on the hyperparameters choice and the signal nature.}
    \label{fig:bdt-nonres}
\end{figure}

The four rows in Figure~\ref{fig:bdt-all} report four groups of models, each defined by a distinct choice of BDT hyperparameters. The three columns correspond to three different fractions of injected signal ($\rm N(S)/N(B)=8.2 \cdot10^{-4},\, 2.1 \cdot 10^{-3},\, 2.7\cdot 10^{-3}$)\footnote{To ease the comparison with previous literature on resonant anomaly detection with the LHCO dataset, we choose the values of signal injection following~\cite{curtains,curtainsf4f}.}. The cut-and-count test is performed for each BDT model for four different values of the anomaly selection threshold ($\rm thr=0.8, \,0.9,\,0.95,\,0.99$) resulting in four different power curves (red, orange, yellow, and green respectively). We observe substantial variance of performances within each group due to the value of $\rm thr$. 
Selecting the best model among the one showed based on a specific signal injection scenario is not a good approach, since this is most likely missing other signals. In fact, there doesn't exist a unique model that is optimal for all signal models. 
This is even more evident if we consider multiple signal benchmarks, as in Figure~\ref{fig:bdt-nonres}. 
In this figure we consider three tests corresponding to the following three hyperparameters sets: $(n_{\rm leaf}=100, \lambda=10^{-2},\, {\rm thr}=0.8)$, $(n_{\rm leaf}=31, \lambda=10^{-1},\, {\rm thr}=0.99)$, and $(n_{\rm leaf}=100, \lambda=10^{-1},\, {\rm thr}=0.99)$. We report their power curves for two signal benchmarks, the resonant one from the LHCO dataset (top row) and the non-resonant one from the RODEM dataset (bottom row). The first BDT model (red line) achieves the highest performances at low signal injection for the resonant signal (top left panel) but performs poorly in all other scenarios; conversely, the second model (orange line) achieves the highest performance at low signal injection for the non-resonant signal but performs poorly in the resonant signal scenarios. Finally, the third model achieves the best performances at high signal injection for both resonant and non-resonant signals but fails at detecting low injection of the resonant signal. 

\subsection{NPLM-classifier for rare anomaly selection}\label{subsec:NPLMclassifier}
\begin{figure}[t!]
    \centering
    \includegraphics[width=1\linewidth]{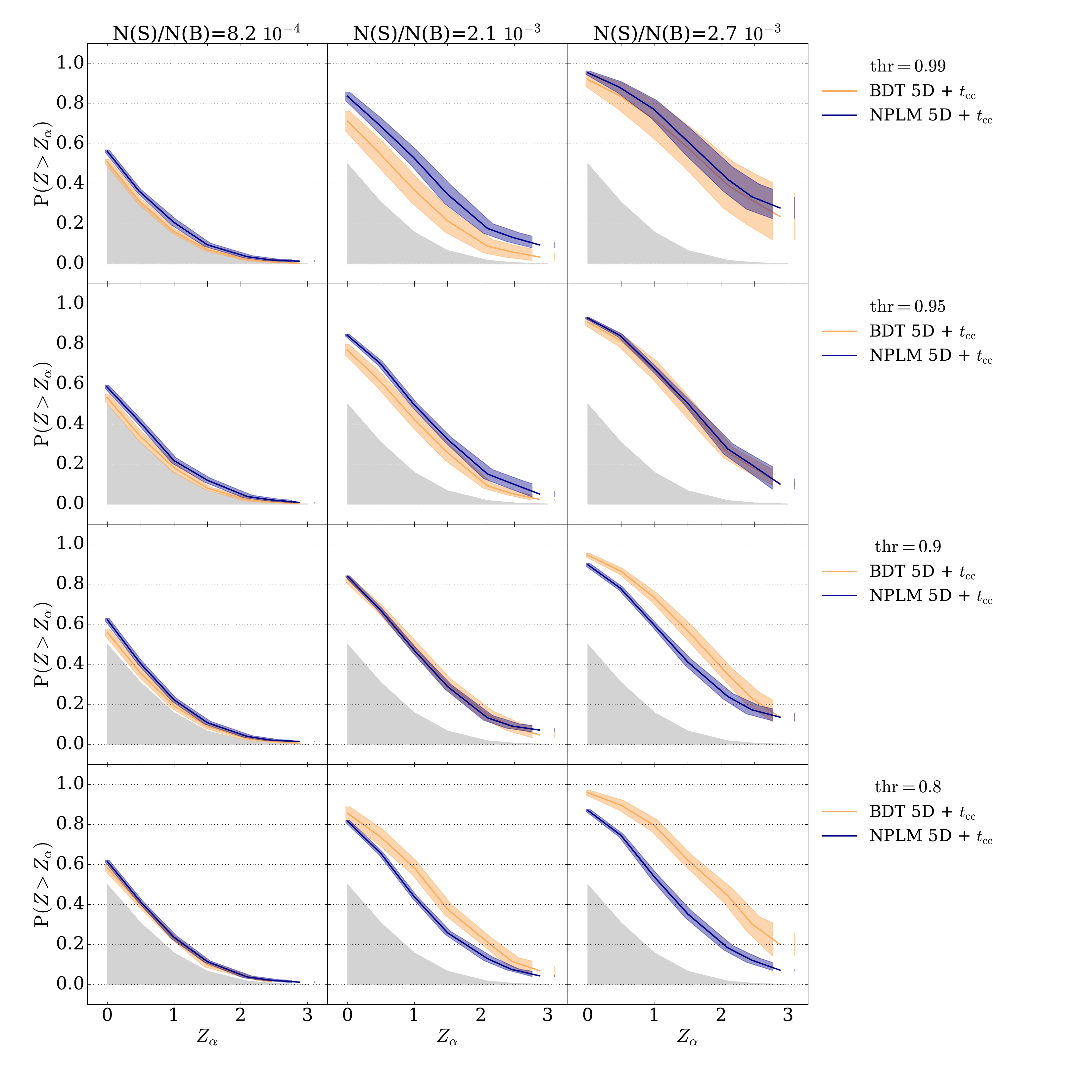}
    \caption{\textbf{BDT-classifier vs. NPLM-classifier at different anomaly selection thresholds.} The panels report the average power of the NPLM based (blue line) and BDT based (orange line) cut-and-count test ($t_{\rm cc}$); each row corresponds to a different value of the selection threshold (from the bottom to the top $\rm thr=0.8,\,0.9,\,0.95,\,0.99$); each column corresponds to a different fraction of signal injection (from the left $\rm N(S)/N(B)=8.2\cdot10^{-4},\, 2.1\cdot 10^{-3},\, 2.7\cdot 10^{-3} $). The shaded bands represent the standard deviation due to the remaining hyperparameters.}
    \label{fig:SsqrtB-thr}
\end{figure}

As an alternative to standard classification, we replace the BDT model with the NPLM model, characterized by a different training strategy. Unlike BDTs, which use tree methods and k-folding for out-of-sample evaluation, NPLM relies on kernel methods trained on the full dataset with in-sample evaluation. This in-sample approach maintains sensitivity to rare signals by avoiding dataset splits, and intentionally encourages overfitting, which is advantageous for anomaly detection within a specific dataset. Consequently, NPLM is expected to outperform BDTs, in low signal injection scenario. We empirically test this hypothesis, as shown in Figure~\ref{fig:SsqrtB-thr}, which compares the average power curve of the cut-and-count test statistic for NPLM classifiers (blue) and BDT classifiers (orange) across different thresholds values, shown separately in the four rows. The shaded bands represent the standard deviation among different hyperparameters choices.
The NPLM models obtains systematically higher performances as the amount of signal injection is reduced. 
At low signal injection ($\rm N(S)/N(B)=8.2 \cdot 10^{-4}$) the NPLM-classifier is better than, or at least equivalent to, the standard BDT-based classifier for any value of $\rm thr$. 
Our empirical observations seem to confirm the hypothesis that in-sample training and testing is beneficial for recovering low signal injections. 

Moreover, by comparing the rows in Figure~\ref{fig:SsqrtB-thr} we notice that the anomaly selection threshold 
is the main source of variance in performances. 
Two possible solutions to mitigate the impact of the selection threshold are presented in the next section.

\subsection{Mitigating the effect of the anomaly selection threshold}
\begin{figure}[t!]
    \centering
    \includegraphics[width=\linewidth]{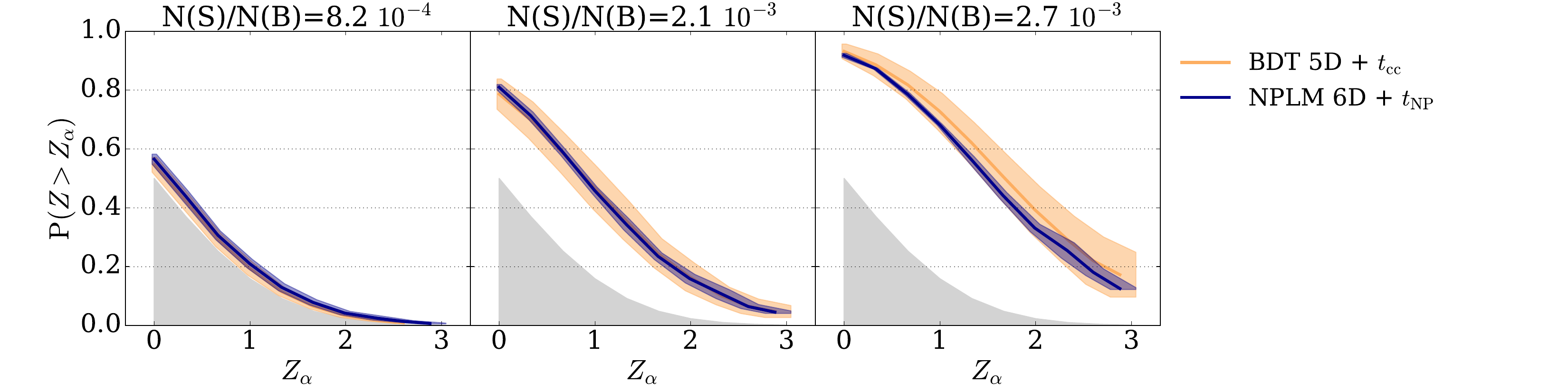}
    \caption{\textbf{NPLM-end-to-end vs. BDT-classifier.} Discovery powers of the NPLM-end-to-end (blue) and the BDT-classifier with cut-and-count test (orange). For both cases, we report the average power and corresponding standard error (shaded bands) over multiple hyperparameters choices. On average, NPLM performs as well as BDT-based models, with significantly reduced variance.}
    \label{fig:NPLM-BDT-variance}
\end{figure}

\paragraph{NPLM-end-to-end.} Using NPLM as an end-to-end approach to simultaneously detect anomalies and perform the test overcomes the problem of variance induced by $\rm thr$. In this case, both the auxiliary variables and the resonant variable are given as an input to the NPLM model and the Neyman-Pearson test statistic is computed end-to-end without introducing a selection threshold.
Figure~\ref{fig:NPLM-BDT-variance} shows the power of the NPLM test run over all six variables compared to the cut-and-count test performed downstream various selection choices over BDT models. The average power over various BDT based approaches is shown in orange, while the NPLM-based one is shown in blue. NPLM performs as well as the average BDT model but with a significantly lower variance.

However, this solution requires a perfect modelling of the background template, which most of the time is not available.
 \begin{figure}[t!]
    \centering
    \includegraphics[width=\linewidth]{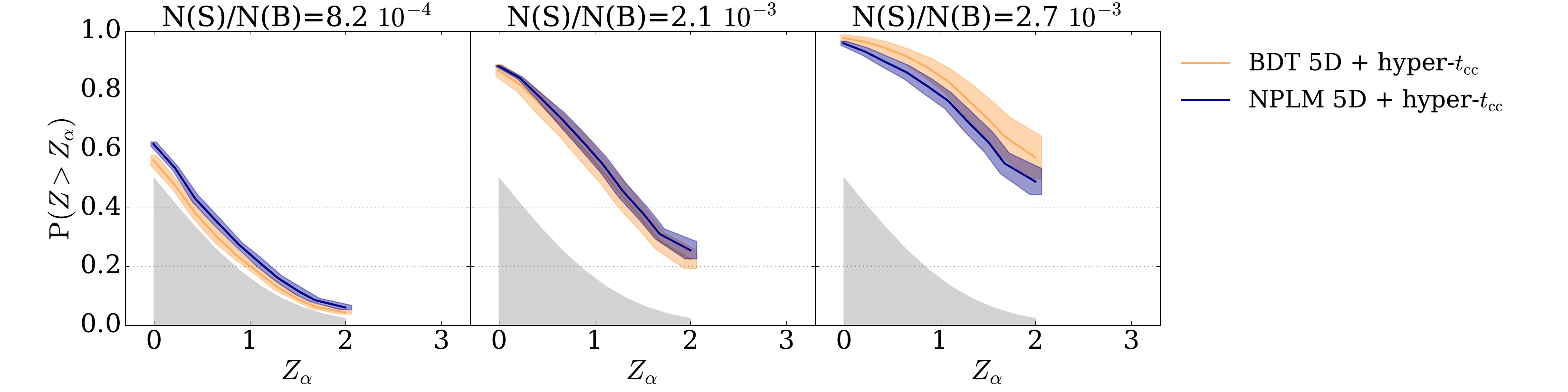}
    \caption{\textbf{NPLM-classifiers vs. BDT-classifiers: hyper-test across anomaly selection thresholds.} Discovery power of the hyper cut-and-count test (hyper-$t_{\rm cc}$). For both the NPLM-classifier (blue) and the BDT-classifier (orange) we report the average power and corresponding standard error (shaded bands) over multiple hyperparameters choice.}
    \label{fig:SsqrtB-all-vs-minp}
\end{figure}

\paragraph{Multiple testing over the anomaly selection threshold.} 
In the absence of an accurate background template, applying a selection and then performing a cut-and-count test remains a viable option. NPLM can be used as a classifier as shown in Section~\ref{subsec:NPLMclassifier}.
To mitigate the impact of the anomaly selection threshold on the cut-and-count test robustness, a multiple test strategy similar to the one proposed in~\cite{Grosso:2024wjt} can be adopted. More precisely, multiple tests based on different threshold values can be computed and the minimum $p$-value can be selected among them, thus building a hyper-test. 
Figure~\ref{fig:SsqrtB-all-vs-minp} shows the sensitivity of such hyper test computed on NPLM-classifiers (blue lines) opposed to BDT-classifiers (orange lines). Calibration is performed empirically by running pseudo-experiments without signal injection (null hypothesis) and computing the hyper-test over them. Beside reducing the variance issue,
we clearly see once more the benefit of using the NPLM approach for rare signal detection.
\begin{table}[H]
    \centering
    \begin{tabular}{cccc}
    \toprule
     \multirow{2}*{statistical strategy} & \multicolumn{3}{c}{N(S)/N(R)}\\
         & $8.2\cdot 10^{-4}$ & $2.1\cdot 10^{-3}$ & $2.7\cdot 10^-3$\\
     \midrule
       BDT 5D + $t_{\rm cc}$        & 0.1 $\pm$ 0.1 & 0.9 $\pm$ 0.3 & 1.7 $\pm$ 0.4\\
       NPLM 6D + $t_{\rm NP}$       & 0.18 $\pm$ 0.06 & 0.96 $\pm$ 0.03 & 1.49 $\pm$ 0.04\\
       BDT 5D + hyper-$t_{\rm cc}$  &0.17 $\pm$ 0.04  &1.2 $\pm$ 0.2 &\textbf{2.4 $\pm$ 0.5}\\
       NPLM 5D + hyper-$t_{\rm cc}$ &\textbf{0.30 $\pm$ 0.04}  &\textbf{1.2 $\pm$ 0.1} &2.0 $\pm$ 0.2\\
    \bottomrule
    \end{tabular}

    \vspace{0.2cm}
    \caption{\textbf{Summary of performances.} Median $Z$-score of the four main approaches considered in this work. The reported values represent the average $Z$-score among different hyperparameters choice and corresponding standard deviation.}
    \label{tab:summary}
\end{table}
Table~\ref{tab:summary} summarizes the performances of the four main approaches studied in this work: the use of a typical BDT-classifier with a cut-and-count test (first row), the use of the default NPLM end-to-end in 6D (second row), the use of a BDT-classifier with hyper-test hyper-$t_{\rm cc}$ designed by performing cut-and-count over multiple values of the selection threshold (third row), and finally an NPLM-classifier with same hyper-$t_{\rm cc}$ test.
For each approach, we report the median $Z$-score among different values of the tunable hyperparameters and corresponding standard error. 
NPLM 6D exhibits significant lower variance than BDT-based approaches, due to the absence of the $\rm thr$ hyperparameter. 
The use of the hyper-$t_{\rm cc}$ test further improves the performances of the classifier based approach while keeping the variance low.
NPLM 5D with hyper-$t_{\rm cc}$ achieves the best performances at low signal injection. For larger signal injection, BDT 5D with hyper-$t_{\rm cc}$ slightly surpasses the NPLM version but remains within one standard deviation from the latter.   

In conclusion, our results suggest that (1) the NPLM training scheme is preferable over the standard classification training scheme for detection of rare signals; (2) the hyper-$t_{\rm cc}$ test is a preferable solution for a robust detection.\\
Several techniques have been developed to adjust the background estimation using control regions in the mass variable~\cite{anode,cathode,LaCathode,curtains,curtainsf4f,drapes,ranode,conrad}, and NPLM could be used as a classifier downstream the output of such methods.

It should be noted that in the BDT case the multiple-testing approach could be further extended to multiple hyperparameters choices as done for NPLM following~\cite{Grosso:2024wjt}. This would potentially reduce the variance of the observed performances. However, additional studies are needed to define an efficient set of BDT models to run multiple test over and to assess the computing load. We leave this for future work.


\subsection{Computing and time resources}
The experiments based on BDTs rely on the Histogram-Based Gradient Boosting library of \texttt{scikit-learn} \cite{scikit-learn} and were executed on CPUs. The average time for a single experiment is $5$ minutes. The experiments based on NPLM were implemented using kernel methods using the \textsc{Falkon}~\cite{falkonlibrary2020,falkonhopt2022} library and were executed on GPUs. The execution time for a single experiment is on average $2$ minutes, comprising of five trainings with different kernel width $\sigma$ combined according to~\cite{Grosso:2024wjt}. 

%% file: tex/outlook.tex
In this paper, we investigated the benefits of integrating the NPLM method in the resonant anomaly detection analysis strategy based on enhanced bump-hunting.
We showed that NPLM can lead to higher detection performances for rare signals, a very relevant regime for realistic applications on LHC data. We argue that the reason for such gain resides in the in-sample nature of the test, opposed to the k-fold approach adopted in standard classification strategies.

We also highlight the problem of high variance due to the choice of hyperparameters in classifier-based resonant anomaly detection. The lack of a priori knowledge about the shape and amount of signal makes the standard hyperparameters tuning procedure not possible, exposing the detection strategy to unforeseen failure modes. 
In particular, we stress the critical role of the selection threshold ($\rm thr$) as the main source of variance in the detection power of the method, and we study two solutions to mitigate such problem based on the NPLM method: 
(1) in presence of a good background modelling running the NPLM algorithm as an end-to-end approach; (2) in absence of a good background modelling, 
using NPLM as a classifier for the anomaly selection stage, and performing a hyper-test over multiple values of $\rm thr$.

The version of NPLM used in this work exploits multiple testing over relevant hyperparameters of the model, reducing the variance of performances due to hyperparameters choice~\cite{Grosso:2024wjt}. Moreover, it does not require a selection step, as the statistical test is computed end-to-end in the training process. For these reasons, the NPLM approach exhibits significantly lower variance in performances. 

However, the end-to-end standard implementation of NPLM can only be applied when the template background is accurately known or the inaccuracy well described by nuisance parameters~\cite{dAgnolo:2021aun}. In real cases this is not always the case, and performing the selection step followed by a background calibration using the sidebands of the resonant variable is a more suitable solution. In this case, we show that the NPLM model can be used as a classifier and the in-sample nature of the test makes the training strategy more sensitive to low signal injection than classifiers based on standard training strategies, like k-folding.
In this case, to further mitigate the impact of the anomaly selection threshold, we perform multiple tests with different values of $\rm thr$ and combine them in a hyper-test which selects the minimum $p$-value~\cite{Grosso:2024wjt}. The resulting test exhibits significantly lower variance, and sensitivity comparable to one of the best tests among the combined ones, improving the median $Z$-score at low signal injection up to a factor 3. These promising results motivate future work to combine NPLM to existing strategies for background estimation like the ones developed in~\cite{anode,cathode,LaCathode,curtains,curtainsf4f,drapes,ranode,conrad}.

In conclusion, we showed two ways to improve the variance and sensitivity of classifier-based resonant anomaly detection using the NPLM method: (1) using NPLM end-to-end, or (2) using NPLM as a classifier and compute a hyper-$t_{\rm cc}$ test. With this work we hope to further engage the community on developing new strategies or refining existing ones that, beside improving on the sensitivity to existing signal benchmarks, aim at a robust response over unpredictable signal variability.

%% file: includes/acknowledgement.tex
The authors would like to acknowledge funding through the SNSF Sinergia grant CRSII5\_193716 ``Robust Deep Density Models for High-Energy Particle Physics and Solar Flare Analysis (RODEM)'', the SNSF project grant 200020\_212127 ``At the two upgrade frontiers: machine learning and the ITk Pixel detector'', and the National Science Foundation under Cooperative Agreement PHY-2019786 (The NSF AI Institute for Artificial Intelligence and Fundamental Interactions, http://iaifi.org/).  Computations in this paper were run on the FASRC Cannon cluster supported by the FAS Division of Science Research Computing Group at Harvard University, and at the University of Geneva using Baobab HPC service.

%% file: main.bbl
\begin{thebibliography}{10}

\bibitem{anode}
Benjamin Nachman and David Shih.
\newblock {Anomaly Detection with Density Estimation}.
\newblock {\em Phys. Rev. D}, 101:075042, 2020, 2001.04990.

\bibitem{cathode}
Anna Hallin, Joshua Isaacson, Gregor Kasieczka, Claudius Krause, Benjamin Nachman, Tobias Quadfasel, Matthias Schlaffer, David Shih, and Manuel Sommerhalder.
\newblock {Classifying anomalies through outer density estimation}.
\newblock {\em Phys. Rev. D}, 106(5):055006, 2022.

\bibitem{LaCathode}
Anna Hallin, Gregor Kasieczka, Tobias Quadfasel, David Shih, and Manuel Sommerhalder.
\newblock {Resonant anomaly detection without background sculpting}.
\newblock 10 2022, 2210.14924.

\bibitem{curtains}
John~Andrew Raine, Samuel Klein, Debajyoti Sengupta, and Tobias Golling.
\newblock {CURTAINs for your sliding window: Constructing unobserved regions by transforming adjacent intervals}.
\newblock {\em Front. Big Data}, 6:899345, 2023, 2203.09470.

\bibitem{curtainsf4f}
Debajyoti Sengupta, Samuel Klein, John~Andrew Raine, and Tobias Golling.
\newblock {CURTAINs Flows For Flows: Constructing Unobserved Regions with Maximum Likelihood Estimation}.
\newblock 5 2023, 2305.04646.

\bibitem{drapes}
Debajyoti Sengupta, Matthew Leigh, John Raine, Samuel Klein, and Tobias Golling.
\newblock Improving new physics searches with diffusion models for event observables and jet constituents.
\newblock {\em Journal of High Energy Physics}, 2024, 04 2024.

\bibitem{ranode}
Ranit Das, Gregor Kasieczka, and David Shih.
\newblock Residual anode, 2023, 2312.11629.

\bibitem{conrad}
Gregor Kasieczka, John~Andrew Raine, David Shih, and Aman Upadhyay.
\newblock Complete optimal non-resonant anomaly detection, 2024, 2404.07258.

\bibitem{CMS:2024lwn}
{Model-agnostic search for dijet resonances with anomalous jet substructure in proton-proton collisions at $\sqrt{s}$ = 13 TeV}.
\newblock 2024.

\bibitem{ATLAS:2020iwa}
{ATLAS Collaboration}.
\newblock {Dijet resonance search with weak supervision using $\sqrt{s}=13$ TeV $pp$ collisions in the ATLAS detector}.
\newblock {\em Phys. Rev. Lett.}, 125(13):131801, 2020, 2005.02983.

\bibitem{nplm}
Gaia Grosso, Marco Letizia, Maurizio Pierini, and Andrea Wulzer.
\newblock Goodness of fit by neyman-pearson testing.
\newblock {\em SciPost Physics}, 16(5), May 2024.

\bibitem{cwola}
Eric~M. Metodiev, Benjamin Nachman, and Jesse Thaler.
\newblock {Classification without labels: Learning from mixed samples in high energy physics}.
\newblock {\em JHEP}, 10:174, 2017, 1708.02949.

\bibitem{cwolabump}
Jack~H. Collins, Kiel Howe, and Benjamin Nachman.
\newblock {Extending the search for new resonances with machine learning}.
\newblock {\em Phys. Rev. D}, 99(1):014038, 2019, 1902.02634.

\bibitem{backtoroots}
Thorben Finke, Marie Hein, Gregor Kasieczka, Michael Kr\"amer, Alexander M\"uck, Parada Prangchaikul, Tobias Quadfasel, David Shih, and Manuel Sommerhalder.
\newblock {Back To The Roots: Tree-Based Algorithms for Weakly Supervised Anomaly Detection}.
\newblock 9 2023, 2309.13111.

\bibitem{DAgnolo:2018cun}
Raffaele~Tito D'Agnolo and Andrea Wulzer.
\newblock {Learning New Physics from a Machine}.
\newblock {\em Phys. Rev. D}, 99(1):015014, 2019, 1806.02350.

\bibitem{DAgnolo:2019vbw}
Raffaele~Tito D'Agnolo, Gaia Grosso, Maurizio Pierini, Andrea Wulzer, and Marco Zanetti.
\newblock {Learning multivariate new physics}.
\newblock {\em Eur. Phys. J. C}, 81(1):89, 2021, 1912.12155.

\bibitem{Letizia:2022xbe}
Marco Letizia, Gianvito Losapio, Marco Rando, Gaia Grosso, Andrea Wulzer, Maurizio Pierini, Marco Zanetti, and Lorenzo Rosasco.
\newblock {Learning new physics efficiently with nonparametric methods}.
\newblock {\em Eur. Phys. J. C}, 82(10):879, 2022, 2204.02317.

\bibitem{Neyman:1933wgr}
Jerzy Neyman, Egon~Sharpe Pearson, and Karl Pearson.
\newblock Ix. on the problem of the most efficient tests of statistical hypotheses.
\newblock {\em Philosophical Transactions of the Royal Society of London. Series A, Containing Papers of a Mathematical or Physical Character}, 231(694-706):289--337, 1933, https://royalsocietypublishing.org/doi/pdf/10.1098/rsta.1933.0009.

\bibitem{dAgnolo:2021aun}
Raffaele~Tito d'Agnolo, Gaia Grosso, Maurizio Pierini, Andrea Wulzer, and Marco Zanetti.
\newblock {Learning new physics from an imperfect machine}.
\newblock {\em Eur. Phys. J. C}, 82(3):275, 2022, 2111.13633.

\bibitem{rudi2017falkon}
Alessandro Rudi, Luigi Carratino, and Lorenzo Rosasco.
\newblock Falkon: An optimal large scale kernel method.
\newblock {\em Advances in neural information processing systems}, 30, 2017.

\bibitem{meanti2020kernel}
Giacomo Meanti, Luigi Carratino, Lorenzo Rosasco, and Alessandro Rudi.
\newblock Kernel methods through the roof: Handling billions of points efficiently.
\newblock In H.~Larochelle, M.~Ranzato, R.~Hadsell, M.F. Balcan, and H.~Lin, editors, {\em Advances in Neural Information Processing Systems}, volume~33, pages 14410--14422. Curran Associates, Inc., 2020.

\bibitem{Grosso:2023hew}
Gaia Grosso.
\newblock {\em {Searching for unexpected New Physics at the LHC with Machine Learning}}.
\newblock PhD thesis, U. Padua (main), Padua U., 1 2023.

\bibitem{Grosso:2024wjt}
Gaia Grosso and Marco Letizia.
\newblock {Multiple testing for signal-agnostic searches of new physics with machine learning}.
\newblock 8 2024, 2408.12296.

\bibitem{LHCOlympics}
Gregor Kasieczka, Benjamin Nachman, and David Shih.
\newblock {Official Datasets for LHC Olympics 2020 Anomaly Detection Challenge (Version v6)}, 2019.

\bibitem{pythia}
Torbjörn Sjöstrand, Stefan Ask, Jesper~R. Christiansen, Richard Corke, Nishita Desai, Philip Ilten, Stephen Mrenna, Stefan Prestel, Christine~O. Rasmussen, and Peter~Z. Skands.
\newblock An introduction to pythia 8.2.
\newblock {\em Computer Physics Communications}, 191:159–177, June 2015.

\bibitem{deFavereau:2013fsa}
J.~de~Favereau, C.~Delaere, P.~Demin, A.~Giammanco, V.~Lema\^\i{}tre, A.~Mertens, and M.~Selvaggi.
\newblock {DELPHES 3, A modular framework for fast simulation of a generic collider experiment}.
\newblock {\em JHEP}, 02:057, 2014, 1307.6346.

\bibitem{Cacciari:2008gp}
Matteo Cacciari, Gavin~P Salam, and Gregory Soyez.
\newblock The anti-ktjet clustering algorithm.
\newblock {\em Journal of High Energy Physics}, 2008(04):063–063, April 2008.

\bibitem{Cacciari:2011ma}
Matteo Cacciari, Gavin~P. Salam, and Gregory Soyez.
\newblock {FastJet User Manual}.
\newblock {\em Eur. Phys. J. C}, 72:1896, 2012, {1111.6097}.

\bibitem{rodemjets}
Knut Zoch, John~Andrew Raine, Debajyoti Sengupta, and Tobias Golling.
\newblock Rodem jet datasets, 2024, 2408.11616.

\bibitem{feynrules}
Adam Alloul, Neil~D. Christensen, Céline Degrande, Claude Duhr, and Benjamin Fuks.
\newblock Feynrules  2.0 — a complete toolbox for tree-level phenomenology.
\newblock {\em Computer Physics Communications}, 185(8):2250–2300, August 2014.

\bibitem{nsubjettiness}
Jesse Thaler and Ken Van~Tilburg.
\newblock Identifying boosted objects with n-subjettiness.
\newblock {\em Journal of High Energy Physics}, 2011(3), March 2011.

\bibitem{scikit-learn}
F.~Pedregosa, G.~Varoquaux, A.~Gramfort, V.~Michel, B.~Thirion, O.~Grisel, M.~Blondel, P.~Prettenhofer, R.~Weiss, V.~Dubourg, J.~Vanderplas, A.~Passos, D.~Cournapeau, M.~Brucher, M.~Perrot, and E.~Duchesnay.
\newblock Scikit-learn: Machine learning in {P}ython.
\newblock {\em Journal of Machine Learning Research}, 12:2825--2830, 2011.

\bibitem{falkonlibrary2020}
Giacomo Meanti, Luigi Carratino, Lorenzo Rosasco, and Alessandro Rudi.
\newblock Kernel methods through the roof: handling billions of points efficiently.
\newblock In {\em Advances in Neural Information Processing Systems 32}, 2020.

\bibitem{falkonhopt2022}
Giacomo Meanti, Luigi Carratino, Ernesto De~Vito, and Lorenzo Rosasco.
\newblock Efficient hyperparameter tuning for large scale kernel ridge regression.
\newblock In {\em Proceedings of The 25th International Conference on Artificial Intelligence and Statistics}, 2022.

\end{thebibliography}
